# Can entropy be defined for and the Second Law applied to the entire universe?

**Arieh Ben-Naim**

**Department of Physical Chemistry**

**The Hebrew University of Jerusalem**

**Givat Ram, Jerusalem 91904**

**Israel**

**Email: ariehbennaim@gmail.com**

 **Abstract**

This article provides answers to the two questions posed in the title. It is argued that, contrary to many statements made in the literature, neither entropy, nor the Second Law may be used for the entire universe. The origin of this misuse of entropy and the second law may be traced back to Clausius himself. More resent (erroneous) "justification" is also discussed.

Keywords: Entropy, Second Law, Thermodynamics, Universe. Free energy

## 1. Introduction

 This article discusses two distinguishable questions: The first, concerns the possibility of defining entropy, and the second, concerns the applicability of the Second Law to the entire universe. Most writers on thermodynamics intertwine these two questions.[1,2] In fact, most people *define* the Second Law as "the law of increasing entropy."[1,2] This is unfortunately, not true. One can define, interpret, and apply the concept of entropy without ever mentioning the Second Law. By the same token, one can also formulate, interpret and apply the Second Law without mentioning the concept of entropy[3-5].

In section 2, we briefly present the relatively recent definition of entropy based on the Shannon measure of information (SMI),[6-8] as well as few possible formulations of the Second Law. Some more details are provided in Appendix A. We show that entropy is involved in the formulation of the Second Law only for processes occurring in isolated systems. The most general formulation of the Second Law does not depend on, nor does it use the concept of entropy.[3-4]

In section 3, we discuss the first question posed in the title of this article: Can entropy be defined for the entire universe? The answer to this question is a definite, No! We shall provide an argument based on SMI supporting this conclusion.

In section 4, we examine the second question posed in the title; whether or not the Second Law can be applied to the universe. We show that contrary to the opinion of a majority of writers on this subject, the Second Law of Thermodynamics *is not applicable* to the entire universe. This is true to any formulation of the Second Law. We shall also trace the origin of this erroneous application of the Second Law.

In section 5, we critically examine a few quotations from the literature where entropy and the Second Law were applied to the universe.

## 2. Definitions of entropy and formulations of the Second Law

In this section, we shall briefly discuss three definitions of entropy. A more detailed discussion may be found in references 3-5. We shall also present a few formulations of the Second Law of thermodynamics.

### 2.1 Clausius' definition

Clausius did not define entropy itself but changes in entropy for a specific process. When a very small amount of heat $dQ$ is transferred to a system at constant temperature, the entropy change is given by: $dS = dQ/T$ . For our purpose, in this article it is suffice to say that Clausius' definition, together with the third law determines the entropy of a system at *equilibrium*. Clausius' definition cannot be used to extend the definition of entropy to non-equilibrium states.  See Appendix B. This observation is important. As we shall see, any other definition of entropy will be accepted as a valid definition of entropy only when calculations based on that definition agrees with the calculations based on Clausius' definition

### 2.2 Boltzmann's definition

There are essentially two "definitions" of the Boltzmann entropy.[9-18]  The first  is the famous equation:

$$S_B = k_B \log W \,, \qquad\qquad (2.1)$$

Here $k_B$ is the Boltzmann constant, and $W$ is the number of accessible microstates of a system having a fixed energy, volume, and number of particles. This equation is based on the assumption that all the microstates of the system are equally probable. From this definition of entropy,  Gibbs derived an expression of the entropy for systems described by other macroscopic variables such as: $(T, V, N), (T, P, N)$ or $(T, V, \mu)$. All these entropies have the same *formal* form[18]

$$S_G = -k_B \sum_i p_i \log p_i \qquad\qquad (2.2)$$

Where $p_i$ are the equilibrium probabilities of finding the system with fixed energy (in a $T, V, N$ ensemble), fixed energy and volume (in a $T, P, N$ ensemble), or fixed energy and number of particles (in a $T, V, \mu$ ensemble, where $\mu$ is the chemical potential).

It is clear that both $S_B$ in (2.1) and $S_G$ in (2.2) are time independent. Also, for any system for which one can calculate the changes in entropy from either (2.1) or (2.2), [for instance, for an expansion of an ideal gas, or mixing of ideal gases], one finds agreement between these calculations and calculations based on Clausius' definition. We emphasize again that all these calculations of entropy changes pertain to equilibrium systems, see also section 3 below.

The second, so-called Boltzmann entropy is based on Boltzmann's definition of the H-function[17]:

$$H(t) = \int f(v,t) \log f(v,t) dv \qquad (2.3)$$

This function is obviously a function of time, and is defined for any distribution of velocities $f(v,t)$. Boltzmann identified the function $-H(t)$ with entropy. This identification is prevalent in the literature even today[16,17,19].

In the following, we shall see that this identification is not valid, mainly because it cannot be compared with calculations based on Clausius' definition. We shall return to this misidentification in the next section.

### 2.3 Definition of entropy based on Shannon's measure of information (SMI)

In this section, we present very briefly the third definition of entropy based on the SMI. More details are provided in Appendix A and in references: 3-8,20.

We start with the SMI, as defined by Shannon[6]. The SMI is defined for any given probability distribution $p_1, p_2, \ldots, p_N$ by

$$H = -K \sum p_i \log p_i \qquad (2.4)$$

Here $K$ is a positive constant and the logarithm was originally taken to the base 2. Here, we use the natural logarithm and include into $K$ the conversion factor between any two bases of the logarithm.

Clearly, the SMI defined in eq. (2.4) is a very general quantity. It is defined for any probability distribution; it can be the probabilities of Head and Tail for tossing a coin, or the six outcomes of a die. It is unfortunate that because of the formal resemblance of (2.4) to Gibbs entropy, the SMI is also referred to as entropy. This has caused a great confusion in both information theory and thermodynamics. This confusion was already recognized by Jaynes who initially adopted the term "entropy" for the SMI, but later realized the potential confusion it could create[3,4,21-25].

In order to obtain the thermodynamic entropy $S$ from the SMI we have to start with the SMI and proceed in two steps: First, apply the SMI to the probability distribution of *locations* and *momenta* of a system of many particles. Second, calculate the maximum of the SMI over all possible such distributions. For more details, see Appendix A. We note here that in this derivation we use the *continuous* analogue of the SMI written as:

$$\text{SMI}(locations\ and\ velocities) = -K \int f(\boldsymbol{R}, \boldsymbol{v}, t) \log f(\boldsymbol{R}, \boldsymbol{v}, t) d\boldsymbol{R} d\boldsymbol{v} \qquad (2.5)$$

However, in actual applications for thermodynamics we always use the discrete definition of the SMI as shown in (2.4). See Appendix A.

The procedure of calculating the distribution that maximizes the SMI in (2.5) is known as the MaxEnt (maxium entropy) principle. We shall refer to this procedure as the MaxSMI, and not as MaxEnt[25].

For a system of non-interacting particles the distribution of locations is uniform and that of velocities (or momenta) is Maxwell Boltzmann. This was originally shown by Shannon himself and some further details are provided in Appendix A and in references 3,4.

After finding the distribution that maximizes SMI in (2.5), denoted by $f^*(\boldsymbol{R}, \boldsymbol{v})$, we can calculate the maximum SMI for that particular distribution, i.e.

$$\text{MaxSMI} = -K \int f^*(\boldsymbol{R}, \boldsymbol{v}) \log f^*(\boldsymbol{R}, \boldsymbol{v}) d\boldsymbol{R} d\boldsymbol{v} \qquad (2.6)$$

Once we calculate the MaxSMI for an ideal gas, we find that the value of the MaxSMI is the same as the entropy of an ideal gas as calculated by Sackur and Tetrode[26,27], which is the entropy of an ideal gas at a specified $E$, $V$ and $N$ at equilibrium. Therefore, we can *define* the entropy of an ideal gas, up to a multiplicative factor $K$, as the MaxSMI as defined in (2.6). Note that unlike the distribution $f(\boldsymbol{R}, \boldsymbol{v}, t)$ in eq. (2.5) the distribution which maximizes the SMI, denote $f^*(\boldsymbol{R}, \boldsymbol{v})$ is not a function of time.

Furthermore, since we know that the locational distribution of an ideal gas at equilibrium is the uniform distribution, and that the velocity distribution of an ideal gas at equilibrium is the Maxwell-Boltzmann distribution, we can identify the distribution $f^*$ in (2.6), which *maximizes* the SMI, with the *equilibrium* distribution $f^{eq}(\boldsymbol{R}, \boldsymbol{v})$, and instead of (2.6) we write

$$\text{MaxSMI} = -\int f^{eq}(\boldsymbol{R}, \boldsymbol{v}) \log f^{eq}(\boldsymbol{R}, \boldsymbol{v}) d\boldsymbol{R} d\boldsymbol{v} \qquad (2.7)$$

Clearly, this MaxSMI *is not* a *function* of *time*. Thus, this definition of the entropy of an ideal gas is equivalent to the Boltzmann entropy, as well as Clausius' entropy. Equivalent, in the sense that calculations of entropy changes between two *equilibrium states* of an ideal gas, based on this definition agree with the result based on Clausius' entropy.

One can also extend the definition of entropy based on the MaxSMI, to a system of interacting particles[20,25]. In which case, the interactions between the particles produce *correlations*, which in turn can be cast in the form of *mutual* information between the particles[3-5,28].

Thus, the procedure of MaxSMI also provides a definition of entropy for systems consisting of interacting particles at equilibrium.

## 2.4 The entropy formulation of the Second Law

The entropy formulation of the SL applies only to isolated systems. We shall formulate it for a one component system having $N$ particles. If there are $k$ components, then $N$ is reinterpreted as a vector comprising the numbers ( $N_1, N_2, \ldots, N_k$ ) where $N_i$ is the number of particles of species $i$.

*For any isolated system characterized by the constant values of the quantities* $(E, V, N)$*, at equilibrium, the entropy is maximum over all possible constrained equilibrium states of the same system.*

## 2.5 The probability formulation of the Second Law

Here, we state the SL for an isolated system, Figure 1. We start with a system of $N$ particles in one compartment, where $N$ is of the order of one Avogadro number, about $10^{23}$ particles. We remove the partition and follow the evolution of the system. At any point in time we define the distribution of particles by the pair of numbers $(n, N - n)$. Of course, we do not count the exact number of particles in each compartment $n$, but we can measure the density of particles in each compartment, $\rho_L = n_L/V$ and $\rho_R = n_R/V$, where $n_L$ and $n_R$ are the numbers of particles in the left (L) and right (R) compartments, respectively $(n_L + n_R = N)$. From the measurement of $\rho_L$ and $\rho_R$ we can also calculate the pair of mole fractions: $x_L = n_L/(n_L + n_R) = \rho_L/(\rho_L + \rho_R)$ and $x_R = n_R/(n_L + n_R) = \rho_R/(\rho_L + \rho_R)$, with $x_L + x_R = 1$. The pair of numbers $(x_L, x_R)$ is referred to as the configuration of the system. This pair $(x_L, x_R)$ is also a probability distribution.

After the removal of the partition between the two compartments, Figure 1b, we can ask: what is the probability of finding the system with a particular configuration $(x_L, x_R)$? We denote this probability by $\Pr(x_L, x_R)$. Since both $x_L$ and $\Pr$ are probabilities, we shall refer to the latter as super probability; $\Pr(x_L, x_R)$ is the probability of finding the probability distribution $(x_L, x_R)$. We can now state the SL for this particular system as follows:

Upon the removal of the partition between the two compartments, the probability distribution (or the configuration) will evolve from the initial one $(x_L, x_R) = (1,0)$, (i.e. all particles in the left compartment) to a new equilibrium state characterized by a uniform locational distribution: $(x_L, x_R) = \left(\frac{1}{2}, \frac{1}{2}\right)$. This means that the densities $\rho_L$ and $\rho_R$ are equal (except for negligible deviations), or equivalently the mole fractions $x_L$ and $x_R$ are equal to ½ . We shall never observe any significant deviation from this new equilibrium state, not in our lifetime, and not in the universe's lifetime.

Note that before we removed the partition in Figure 1a, the probability of finding the configuration (1, 0) is one. This is an equilibrium state, and all the particles are, by definition of the initial state, in the L compartment.

We refer to the super probability of finding the configuration $(x_L, x_R)$, denoted by $\Pr(x_L, x_R)$, as the probability of the configuration attained *after* the removal of the partition, when $x_L$ can, in principle, attain any value between zero and one. Therefore, the super probability of obtaining the configuration (1, 0) is negligibly small. On the other hand, the super probability of obtaining the configuration in the neighborhood of $\left(\frac{1}{2}, \frac{1}{2}\right)$ is, for all practical purposes nearly one.[3-5,28] This means that after the removal of the partition, and reaching an equilibrium state,  the ratio of the super probabilities of the initial configuration (1, 0) and the final configuration, i.e. in the neighborhood of $\left(\frac{1}{2}, \frac{1}{2}\right)$, is almost infinity. With $N \approx 10^{23}$, this is an unimaginably large number). Thus, we can say that for $10^{23}$

$$\frac{\Pr(\text{final configuration})}{\Pr(\text{initial configuration})} \approx \text{infinity} \quad (2.8)$$

This is the *probability formulation* of the Second Law for this particular experiment. This law states that starting with an equilibrium state where all particles are in L, and removing the constraint (the partition), the system will evolve to a new equilibrium configuration which has a probability overwhelmingly larger than the initial configuration.

Note carefully that if $N$ is small, then the evolution of the configuration will not be monotonic, and the ratio of the super probabilities in the equation above is not near infinity. For very large $N$, the evolution of the configuration is also not strictly monotonic, and the ratio of the super probabilities is not strictly, infinity. However, in practice, whenever $N$ is large, we shall never observe any deviations from monotonic change of the configuration from the initial value $(1, 0)$ to the final configuration $\left(\frac{1}{2}, \frac{1}{2}\right)$. Once the final equilibrium state is reached [i.e. that the configuration is within experimental error $\left(\frac{1}{2}, \frac{1}{2}\right)$], it will stay there *forever*, or equivalently it will be found with probability one.

The distinction between the strictly mathematical monotonic change and the practical change is important. The process is mathematically always reversible, i.e. the initial state will be visited. However, in practice the process is irreversible; we shall never see the reversal to the initial state.

Let us repeat the probability formulation of the Second Law for the more general process in Figure 2. We start with an initial constrained equilibrium state. We remove the constraint, and the system's configuration will evolve with probability (nearly) one, to a new equilibrium state, and we shall never observe reversal to the initial state. "Never" here, means never in our lifetime, nor in the lifetime of the universe.

Note carefully that we formulated the SL in terms of the probability *without* mentioning entropy. This is in sharp contrast with most formulations of the Second Law.

So far we have formulated the Second Law in terms of probabilities. One can show that this formulation is equivalent to the entropy formulation of the Second Law. The relationship between the two formulations (for isolated systems) is[3-5,28]

$$\frac{\Pr(f)}{\Pr(in)} = \exp\left[\frac{\Delta S(in \to f)}{k_B}\right] \qquad (2.9)$$

Here, the probability ratio on the left hand side is the same as in eq. (2.8), except that here the initial ($in$) and the final ($f$) configurations are more general than in eq. (2.8).

We now very briefly discuss two more generalizations of the Second Law. For a process at $T, V, N$ constants, we can write the relationship

$$\frac{\Pr(f)}{\Pr(in)} = \exp\left[-\frac{\Delta A(in \to f)}{k_B T}\right], \qquad (T, V, \boldsymbol{N}\ system) \qquad (2.10)$$

Here, $\Delta A = \Delta E - T\Delta S$ is the change in the Helmholtz energy of the system for the general process as described in Figure 2, except that the total process is carried out at constant temperature ($T$), rather than constant energy ($E$).

Finally, for a process at constant $P, T, \boldsymbol{N}$ we have

$$\frac{\Pr(f)}{\Pr(in)} = \exp\left[-\frac{\Delta G(in \rightarrow f)}{k_B T}\right], \qquad (T, P, \boldsymbol{N} \; system) \qquad (2.11)$$

Here, $\Delta G = \Delta A + P \Delta V$, and $\Delta V$ is the volume change in the process. Note that if we have a one-component system, and each compartment has the same $P$ and $T$, then the chemical potential of each compartment will also be the same. Therefore, no process will occur in such a system. However, in a multi-component system, there will be processes for which $\Delta G < 0$.

Comparing equations (2.9), (2.10) and (2.11), we see that the thermodynamic formulations of the Second Law are *different* for different systems (entropy maximum for $(E, V, N)$ system, Helmholtz energy minimum for $(T, V, \boldsymbol{N})$ system and Gibbs energy minimum for $(T, P, \boldsymbol{N})$ system). On the other hand, the probability formulation is the same for all systems; the probability ratio on the left hand side of equations (2.9), (2.10) and (2.11) is of the order $e^N$. For $N \approx 10^{23}$, this is practically infinity as noted above.

Clearly, the probability formulation is far more general than any other thermodynamic formulation of the Second Law. Therefore this formulation is preferable.

## 3. Can we define the entropy for the entire universe?

Before we attempt to *define* the entropy of the universe it should be noted that Clausius himself *used* the term entropy to formulate the Second Law as:

### *The entropy of the world always increases*

Clausius, who "invented" the concept of entropy, did not *define* entropy. He also had no idea what entropy means on a molecular level. This understanding came much later. Now that we have a clear-cut understanding of entropy it is also clear-cut that entropy cannot be defined for the entire universe. We examine the three "definitions" of entropy:

(i)      Clausius'definition

Clearly, Clausius' "definition" cannot be applied to the entire universe. First, because the universe is not a well-defined, thermodynamic system at constant temperature for which the quantity $dQ/T$ may be applied.

Second, we do not know whether the universe is finite or infinite. Therefore, we cannot use the integral over $C_p/T$ to calculate the "absolute" entropy based on the third law.

(ii)      Boltzmann's definition

Boltzmann's definition requires the knowledge of the number of accessible microstates, W, of the entire universe. Since we do not know how to describe the universe by the conventional thermodynamic quantities such as volume $V$, energy $E$, and composition $\boldsymbol{N} = (N_1, N_2, \ldots, N_c)$, there is no way we can even imagine calculating the number of microstates.

Although some people talk about the "wave function" of the entire universe, hence also about the "Schrödinger equation" for the entire universe, this is clearly meaningless, as long as we have no idea how many particles are in the universe.

(iii)     The definition based on the SMI.

The most straightforward answer regarding the question posed in the heading of this section is provided by this definition of entropy. This definition requires the knowledge of the distribution of locations and momenta of all particles in the system which maximizes the SMI of the system. Clearly, since we do not know the *volume* of the universe, nor the number of particles in the universe, there is no point in discussing the locational distribution of particles in the universe.

It should be noted that the above comment regarding the impossibility of defining the entropy of the universe applies to a universe viewed as an ideal gas – an assumption made frequently in popular science books. We have seen that even with such an unjustified assumption of the universe, one cannot define its entropy. In reality, the universe is far from being an ideal gas. One must also include the interaction energy among all the particles in the universe in any attempt of estimating its entropy. This adds another argument in favor of the impossibility of defining the entropy of the universe – let alone, giving numerical values of the entropy of the universe in the present, in the early past, or at the Big Bang (see section 5).

Finally, we add one more argument to the impossibility of defining the entropy of the universe. We know that life exists in the universe. We know that life is part of the universe. We also know that entropy cannot be defined for any living system[29]. Therefore, we cannot hope to define the entropy of the universe even if we knew that it is finite, having a volume $V$, total energy $E$, and composition $N$.

## 4.   Is the Second Law applicable to the entire universe?

As we have noted above, Clausius already formulated the Second Law in terms of the ever increasing entropy of the universe.

The prevalent view is that if, in an isolated system the entropy *decreases*, then entropy must increase "somewhere" in the universe. This is of course a meaningless statement, unless one can show *where* and how the entropy increases in the universe. Needless to say, that no one has ever shown *where* in the universe the entropy must increase – and what is the process for which the entropy increases.

In the rest of this section we discuss the erroneous argument leading from the entropy formulation of the Second Law applied to an isolated system, to the entropy formulation of the Second Law for the entire universe.

Let us now examine the reasoning underlying the conclusion that if the entropy of a system decreases, then the entropy of the entire universe must increase

To examine this question we discuss the passage from the entropy formulation of the Second Law for isolated system, to the Helmholtz energy formulation for a $T,V,N$ system.

Consider a closed system having a fixed volume $V$, and constant temperature $T$. Such a system can exchange heat with its surroundings. Normally, we keep the temperature of the *system* fixed by placing it in a heated *bath*. The heat bath is supposed to be very large compared with the system, such that when there is a small exchange of heat between the system and the bath, the bath's temperature is not affected.

Before we formulate the SL for such a system we defined the Helmholtz energy by: $A = E - TS$.

Here we have on the right hand side of the equation the energy $E$, the temperature $T$ and the entropy $S$.

This definition of the Helmholtz energy is valid for any system at equilibrium. In this definition we did not specify the independent variables with which we characterize the system. We have the liberty to choose any set of independent variables we choose, say $(E, V, N), (T, V, N)$ or $(T, P, N)$. However, if we want to formulate the SL in terms of the Helmholtz energy we have no choice but to choose the independent variables $(T, V, N)$. For a $(T, V, N)$ system the Helmholtz energy formulation of the Second Law is:

***For any $(T, V, N)$ system at equilibrium the Helmholtz energy has a minimum over all possible constrained equilibrium states of the same system.***

Recall that the entropy formulation of the SL was valid only for *isolated* systems, i.e. for $(E, V, N)$ systems. The Helmholtz energy formulation is valid only systems that are isothermal (constant $T$), and isochoric (constant $V$), as well as closed (constant $N$).

An equivalent statement of the Helmholtz energy formulation is:

***Removing any constraint from a constrained equilibrium state in a $(T, V, N)$ system will result in a decrease in Helmholtz energy.***

Note carefully that it is only for a process at constant $(T, V, N)$ that this formulation is valid. It is not true that the Helmholtz energy decreases for any process occurring in any thermodynamic system.

The Helmholtz energy formulation is equivalent to the entropy formulation in the following sense.

Mathematically, we start with the *entropy function $S(E, V, N)$* and make a transformation of variables from $(E, V, N)$ to $(T, V, N)$. One can prove that the minimum of the Helmholtz energy *function $A(T, V, N) = E(T, V, N) - TS(T, V, N)$* follows from the maximum entropy. This transformation is known as the Legendre transformation[20]. An excellent exposition of this transformation is provided in Callen's book (1985)[20].

The reverse of the last statement is also true; i.e. from the minimum principle of the Helmholtz energy function one can get the maximum entropy principle provided that we transform back from the $(T, V, N)$ to the $(E, V, N)$ variables.

We present a qualitative argument based on the experiment shown in Figure 3. Here, we have a $(T, V, N)$ system immersed in a very large thermal bath. We allow exchange of heat between the system and the bath, but no exchange of volume (i.e. constant $V$), and no exchange of matter (constant $N$). The combined system plus the bath is isolated, and will be referred to as *total.*

Now suppose we remove a constraint in the equilibrated system (this could be a removal of the partition or adding a catalyst which enables chemical reactions). Suppose that as a result of the process in the system, the entropy of the system has changed, denote this change by $\Delta S(sys)$. This change can be either positive or negative. Also, suppose that the energy of the system changed, and denote this change by $\Delta E(sys)$. Again this change can be either positive or negative.

Since the process was carried out at constant $T$, and since the total energy of the system plus bath is constant, we must have the following equality for the combined system plus bath, i.e., for the *total*

$$\Delta E(total) = \Delta E(sys) + \Delta E(bath) = 0 \qquad (4.1)$$

The system and the bath can exchange only heat (no work). Therefore, the change $\Delta E(sys)$ must be a result of the flow of heat $\Delta Q(bath \rightarrow system)$. If $\Delta E(sys)$ is positive, then $\Delta Q$ is positive (heat flows from the bath into the system). If $\Delta E(sys)$ is negative, then $\Delta Q$ is negative.

Now, we use Clausius' principle. Since the bath exchanges only heat with the system, the entropy change of the bath is $\Delta S(bath) = -\Delta Q/T$. If $\Delta Q$ is positive (i.e. heat flows into the system), then $\Delta S(bath)$ will be *negative*. We still cannot say anything about the sign of the change in the entropy of the system. The reason is that the system's entropy has changed both because of the heat flow and the processes that occurred following the removal of some constraints.

The system, plus bath which we refer to as *total*, is isolated. Therefore, the entropy change for the *total* must be positive, i.e.

$$\Delta S(total) = \Delta S(sys) + \Delta S(bath)$$

$$= \Delta S(sys) - \Delta Q/T = \Delta S(sys) - \Delta E(sys)/T \geq 0 \qquad (4.2)$$

Rearranging this equation we obtain

$$-T\Delta S(total) = \Delta E(sys) - T\Delta S(sys) = \Delta A(sys) \qquad (4.3)$$

This is a remarkable equation. The entropy change of the *total* (system plus bath) times the temperature $T$ is equal to minus the Helmholtz energy change of the *system*. Although we do not know the values of the changes $\Delta E(sys)$ and $\Delta S(sys)$, each can be either positive or negative. We can say that $\Delta S(total)$ is positive (because of the entropy formulation of the SL for the isolated *total*). Therefore, we can conclude that the sign of $\Delta A(sys)$ must be negative. Thus we have shown that the Helmholtz energy formulation, for the $(T, V, N)$ system, follows from the entropy formulation for the $(E, V, N)$ system.

It is absolutely important to remember the assumptions that we made in deriving this conclusion: First, that the process in the system occurred at constant *T*, *V*, *N*. The variables *V* and *N* of the system were kept constant by means of rigid and impermeable walls. The temperature of the system was kept constant by requiring that the bath be so large that even when it exchanges heat $(\Delta Q)$ with the system its temperature is not affected. Second, and most importantly the system + bath is isolated. Therefore, we could apply the entropy formulation to the *total*. From this entropy formulation we derived the Helmholtz energy formulation for the $(T, V, N)$ system. Third, we assume that the bath can only exchange heat with the system. No other processes are presumed to take place in the bath.

In Atkins' book (2007)[2] one can find the same process but reaching the absurd conclusion that:

*"A change in A is just a disguised form of the change in the total entropy of the universe when the temperature and the volume of the system are constant."*

The change in *A* is *not* a disguised form of the change in the total entropy, and the entropy of the universe is *not* defined! Atkin's fatal error is a result of his failure to realize that none of the conditions imposed on the bath holds when we replace the bath by the universe, Figure 4.

The universe is not at a constant temperature. We do not know whether the universe is isolated or not. The entropy of the universe has never been defined, and the Helmholtz energy change of the system is *not* a disguised form of entropy change of the universe (whatever this means).

We have stated the principle of the minimum of the Helmholtz energy with respect to all possible constrained equilibrium states having the same $(T, V, N)$. One can repeat the same argument for a *P, T, N* system. For details see Ben-Naim (2017)[4]

We summarize here the essential assumptions made in passing from the entropy formulation for an isolated system to the Helmholtz energy formulation for a system in a constant temperature bath.

1. That the system and the bath form an isolated system.
2. That the change in the entropy of the bath occurred at constant temperature.
3. That the *only process* occurring in the bath is the flow of heat *dQ* from the system into the bath.

It is only within the assumptions 2 and 3 that one can claim that the entropy of the bath has changed by the amount $dQ/T$. Add to this assumption 1, and one arrives at the conclusion that the Helmholtz energy of the system must have decreased in the spontaneous process.

Obviously, once we replace the bath at constant *T* by the entire universe, all these three assumptions become invalid. Therefore, we can safely conclude that that when a spontaneous process occurring in a system at constant temperature causes a decrease in the entropy of the system, the entropy of the bath must increase and the total entropy of the system, plus the bath will increase. This conclusion cannot be extended for a "bath" replaced by the entire universe.

With this conclusion the most powerful argument used to "justify" the formulation of the Second Law in terms of "entropy of the universe always increases," collapses.

Therefore, it is suggested that neither the concept of entropy, nor the Second Law should be used in connection with the entire universe.

## 5. Examples of misuses of entropy and the Second Law for the entire universe

As we have noted above the formulation of the Second Law as the ever increasing entropy has been featuring in literatures ever since Clausius coined the term "entropy" and (unjustifiably) concluded that the entropy of the universe always increases.

As is well known, Clausius formulated one version of the Second Law (heat flows from a hot body to a cold body). Unfortunately, Clausius failed in *over generalizing* the Second Law. His well-known and well quoted statement:

***"The entropy of the universe always increases."***

I do not know how Clausius arrived at this formulation of the Second Law. One can only guess that what has motivated him to conclude that the entropy of the universe always increases is the erroneous argument described in the previous section. Unfortunately, such a generalization is unwarranted. Although most authors will tell you that the entropy of the universe always increases, the truth is that no one has ever measured or calculated the entropy of the universe. In fact, no one has ever *defined* the entropy of the universe. Therefore, any statement regarding the change in the "entropy of the universe" is meaningless.

Here is a quotation from a relatively recent book by Atkins (2007)[2]:

*"The entropy of the universe increases in the course of any spontaneous change. The key word here is **universe**; it means, as always in thermodynamics, the system together with its surroundings. There is no prohibition of the system or the surroundings **individually** undergoing a decrease in entropy provided that there is a compensating change elsewhere."*

Such a generalization is not only untrue, it is simply meaningless. This is only one example of an unwarranted generalization. Penrose (1989)[30] and others not only discuss the entropy of the universe, but also give *numbers*, stating what the entropy of the universe is at present, how much it was in the early universe, and perhaps also at the Big-Bang.

These are meaningless numbers assigned to a meaningless quantity (the entropy of the universe), to the state of the universe at a highly speculative time (at the Big-Bang). More on this in references 3,4,8.

More recently Carroll[31] goes a step further, not only assigning numbers to the entropy of the universe at present, and not only estimating the entropy of the universe at the speculative event referred to as the Big Bang, but also claiming that this low-entropy "fact" can explain many things such as "why we remember the past but not the Future…"

All these senseless claims could have been averted has Carroll, as well as many others recognize that it is meaningless to talk about the entropy of the universe. It is a fortiori meaningless to talk about the entropy of the universe at some distant point in time.

## Appendix A:  Definition of Entropy bases on Shannon's Measure of Information

In this Appendix we derive the entropy function for an ideal gas. We start with SMI which is definable to any probability distribution. We apply the SMI to two specific molecular distributions; the locational and the momentum distribution of all particles. Next, we calculate the distribution which maximizes the SMI. We refer to this distribution as the *equilibrium* distribution. Finally, we apply two corrections to the SMI, one due to Heisenberg uncertainty principle, the second due to the indistinguishability of the particles. The resulting SMI is, up to a multiplicative constant equal to the entropy of the gas, as calculated by Sackur and Tetrode based on Boltzmann definition of entropy.

In previous publication[3-5], we discussed several advantages to the SMI-based definition of entropy. For our purpose in this article the most important aspect of this definition is the following:

The entropy is *defined* as the maximum value of the SMI. As such, it is not a function of time.

### A.1. The Locational SMI of a Particle in a 1D Box of Length L

Suppose we have a particle confined to a one-dimensional (1D) "box" of length L. Since there are infinite points in which the particle can be within the interval (0, L). The corresponding locational SMI must be infinity. However we can defined, as Shannon did, the following quantity by analogy with the discrete case:

$$H[f] = -\int f(x) \log f(x) dx \qquad (A.1)$$

This quantity might either converge or diverge, but in any case, in practice we shall use only differences of this quantity. It is easy to calculate the density which maximizes the locational SMI, $H(f)$ in (A.1) which is:

$$f_{eq}(x) = \frac{1}{L} \qquad (A.2)$$

The use of the subscript eq (for equilibrium) will be cleared later, and the corresponding SMI calculated by (A.1) is:

$$H_{max}(\text{locations in } 1D) = \log L \qquad (A.3)$$

We acknowledge that the location X of the particle cannot be determined with absolute accuracy, i.e. there exists a small interval, $h_x$ within which we do not care where the particle is. Therefore, we must correct equation (A.3) by subtracting $\log h_x$. Thus, we write instead of (A.3):

$$H(location\ in\ 1D) = \log L - \log h_x \qquad (A.4)$$

We recognize that in (A.4) we effectively defined **SMI** for a finite number of intervals $\boldsymbol{n = L/h}$. Note, that when $\boldsymbol{h_x \to 0}$, $\boldsymbol{H}$ in (A.4) diverges to infinity. Here, we do not take the mathematical limit, but we stop at $\boldsymbol{h_x}$ small enough but not zero. Note also that in writing (A.4) we do not have to specify the units of length, as long as we use the same units for L and $\boldsymbol{h_x}$.

### A.2. The Velocity SMI of a Particle in 1D "Box" of Length L

Next, we calculate the probability distribution that maximizes the continuous SMI, subject to two conditions:

$$\int_{-\infty}^{\infty} f(x)dx = 1 \qquad (A.5)$$

$$\int_{-\infty}^{\infty} x^2 f(x)dx = \sigma^2 = constant \qquad (A.6)$$

The result is the Normal distribution:

$$f_{eq}(x) = \frac{\exp[-x^2/\sigma^2]}{\sqrt{2\pi\sigma^2}} \qquad (A.7)$$

The subscript eq. for equilibrium will be clear later. Applying this result to a classical particle having average kinetic energy $\frac{m<v_x^2>}{2}$, and identifying the standard deviation $\sigma^2$ with the temperature of the system:

$$\sigma^2 = \frac{k_B T}{m} \qquad (A.8)$$

We get the equilibrium velocity distribution of one particle in 1D system:

$$f_{eq}(v_x) = \sqrt{\frac{m}{2mk_B T}} \exp\left[\frac{-mv_x^2}{2k_B T}\right] \qquad (A.9)$$

where $k_B$ is the Boltzmann constant, $m$ is the mass of the particle, and $T$ the absolute temperature. The value of the continuous SMI for this probability density is:

$$H_{max}(velocity\ in\ 1D) = \frac{1}{2}\log(2\pi e k_B T/m) \qquad (A.10)$$

Similarly, we can write the momentum distribution in 1D, by transforming from $v_x \to p_x = mv_x$, to get:

$$f_{eq}(p_x) = \frac{1}{\sqrt{2\pi m k_B T}} \exp\left[\frac{-p_x^2}{2mk_B T}\right] \qquad (A.11)$$

and the corresponding maximal SMI:

$$H_{max}(momentum\ in\ 1\ D) = \frac{1}{2}\log(2\pi e m k_B T) \qquad (A.12)$$

As we have noted in connection with the locational SMI, the quantities (A.11) and (A.12) were calculated using the definition of the *continuous* SMI. Again, recognizing the fact that there is a limit to the accuracy within which we can determine the velocity, or the momentum of the particle, we correct the expression in (A.12) by subtracting $\log h_p$ where $h_p$ is a small, but infinite interval:

$$H_{max}(momentum\ in\ 1D) = \frac{1}{2}\log(2\pi emk_BT) - \log h_p \qquad (A.13)$$

Note again that if we choose the units of $h_p$ (of momentum as: $mass\ length/time$) the same as of $\sqrt{mk_BT}$, then the whole expression under the logarithm will be a pure number.

### A.3. Combining the SMI for the Location and Momentum of one Particle in 1D System

In the previous two sections, we derived the expressions for the locational and the momentum SMI of one particle in 1D system. We now combine the two results. Assuming that the location and the momentum (or velocity) of the particles are independent events we write

$$H_{max}(location\ and\ momentum) = H_{max}(location) + H_{max}(momentum)$$

$$= \log\left[\frac{L\sqrt{2\pi emk_BT}}{h_x h_p}\right] \qquad (A.14)$$

Recall that $h_x$ and $h_p$ were chosen to eliminate the divergence of the SMI for a continuous random variables; location and momentum.

In (A.14) we assume that the location and the momentum of the particle are independent. However, quantum mechanics imposes restriction on the accuracy in determining both the location $x$ and the corresponding momentum $p_x$. In Equations (A.4) and (A.13) $h_x$ and $h_p$ were introduced because we did not care to determine the location and the momentum with an accuracy greater that $h_x$ and $h_p$, respectively. Now, we must acknowledge that nature imposes upon us a limit on the accuracy with which we can determine both the location and the corresponding momentum. Thus, in Equation (A.14), $h_x$ and $h_p$ cannot both be arbitrarily small, but their product must be of the order of Planck constant $h = 6.626 \times 10^{-34}\ J\ s$. Thus we set:

$$h_x h_p \approx h \qquad (A.15)$$

And instead of (A.14), we write:

$$H_{max}(location\ and\ momentum) = \log\left[\frac{L\sqrt{2\pi emk_BT}}{h}\right] \qquad (A.16)$$

### A.4. The SMI of a Particle in a Box of Volume V

We consider again one simple particle in a box of volume $V$. We assume that the location of the particle along the three axes $x$, $y$ and $z$ are independent. Therefore, we can write the SMI of the location of the particle in a cube of edges $L$, and volume $V$ as:

$$H(location\ in\ 3D) = 3H_{max}(location\ in\ 1D) \qquad (A.17)$$

Similarly, for the momentum of the particle we assume that the momentum (or the velocity) along the three axes *x*, *y* and *z* are independent. Hence, we write:

$$H_{max}(momentum\ in\ 3D) = 3H_{max}(momentum\ in\ 1D) \qquad (A.18)$$

We combine the SMI of the locations and momenta of one particle in a box of volume *V*, taking into account the uncertainty principle. The result is

$$H_{max}(location\ and\ momentum\ in\ 3D) = 3\log[\frac{L\sqrt{2\pi e m k_B T}}{h}] \qquad (A.19)$$

## A.5. The SMI of Locations and Momenta of N indistinguishable Particles in a Box of Volume V

The next step is to proceed from one particle in a box to *N* independent particles in a box of volume *V*. Giving the location $(x, y, z)$, and the momentum $(p_x, p_y, p_z)$ of one particle within the box, we say that we know the *microstate* of the particle. If there are *N* particles in the box, and if their microstates are independent, we can write the SMI of *N* such particles simply as *N* times the SMI of one particle, i.e.,

$$\text{SMI}(\text{of } N \text{ independent particles}) = N \times \text{SMI}(\text{one particle}) \qquad (A.20)$$

This Equation would have been correct when the microstates of all the particles where independent. In reality, there are always correlations between the microstates of all the particles; one is due to *intermolecular interactions* between the particles, the second is due to the *indistinguishability* between the particles. We shall discuss these two sources of correlation separately.

*(i) correlation due to indistinguishability*

Recall that the microstate of a single particle includes the location and the momentum of that particle. Let us focus on the location of one particle in a box of volume *V*. We have written the locational SMI as:

$$H_{max}(location) = \log V \qquad (A.21)$$

Recall that this result was obtained for the continuous locational SMI. This result does not take into account the divergence of the limiting procedure. In order to explain the source of the correlation due to indistinguishability, suppose that we divide the volume *V* into a very large number of small cells each of the volume $V/M$. We are not interested in the exact location of each particle, but only in which cell each particle is. The total number of cells is *M*, and we assume that the total number of particles is $N \ll M$. If each cell can contain at most one particle, then there are *M* possibilities to put the first particle in one of the cells, and there are $M - 1$ possibilities to put the second particle in the remaining empty cells. Altogether, we have $M(M - 1)$ possible microstates, or configurations for two particles. The probability that one particle is found in cell *i*, and the second in a different cell *j* is:

$$\Pr(i, j) = \frac{1}{M(M-1)} \qquad (A.22)$$

The probability that a particle is found in cell *i* is:

$$\Pr(j) = \Pr(i) = \frac{1}{M} \qquad (A.23)$$

Therefore, we see that even in this simple example, there is correlation between the events "one particle in $i$" and one particle in $j$":

$$g(i,j) = \frac{\Pr(i,j)}{\Pr(i)\Pr(j)} = \frac{M^2}{M(M-1)} = \frac{1}{1-\frac{1}{M}} \qquad (A.24)$$

Clearly, this correlation can be made as small as we wish, by taking $M \gg 1$ (or in general, $M \gg N$). There is another correlation which we cannot eliminate and is due to the indistinguishability of the particles.

Note that in counting the total number of configurations we have implicitly assumed that the two particles are labeled, say red and blue. In this case we count the two configurations in Figure 5a, as *different* configurations: "blue particle in cell $i$, and red particle in cell $j$," and "blue particle in cell $j$, and red particle in cell $i$."

Atoms and molecules are indistinguishable by nature; we cannot label them. Therefore, the two microstates (or configurations) in Figure 5b are indistinguishable. This means that the total number of configurations is not $M(M-1)$, but:

$$number\ of\ configurations = \frac{M(M-1)}{2} \to \frac{M^2}{2}, \text{for large } M \qquad (A.25)$$

For very large $M$ we have a correlation between the events "particle in $i$" and "particle in $j$":

$$g(i,j) = \frac{\Pr(i,j)}{\Pr(i)\Pr(j)} = \frac{M^2}{M^2/2} = 2 \qquad (A.26)$$

For $N$ particles distributed in $M$ cells, we have a correlation function (For $M \gg N$):

$$g(i_1, i_2, \dots, i_n) = \frac{M^N}{M^N/N!} = N! \qquad (A.27)$$

This means that for $N$ indistinguishable particles we must divide the number of configurations $M^N$ by $N!$. Thus in general by removing the "labels" on the particles the number of configurations is *reduced* by $N!$. For two particles the two configurations shown in Figure 5a reduce to one shown in Figure 5b.

Now that we know that there are correlations between the events "one particle in $i_1$", "one particle in $i_2$" … "one particle in $i_n$", we can define the *mutual information* corresponding to this correlation. We write this as:

$$I(1; 2; \dots; N) = \ln N! \qquad (A.28)$$

The SMI for $N$ particles will be:

$$H(N\ particles) = \sum_{i=1}^{N} H(one\ particle) - \ln N! \qquad (A.29)$$

For the definition of the mutual information, see refs: 3-5.

Using the SMI for the location and momentum of $N$ independent particles in (A.20) we can write the final result for the SMI of $N$ indistinguishable (but non-interacting) particles as:

$$H(N \text{ indistinguishable}) = N \log V \left(\frac{2\pi m e k_B T}{h^2}\right)^{3/2} - \log N! \qquad (A.30)$$

Using the Stirling approximation for $\log N!$ in the form (note again that we use the natural logarithm):

$$\log N! \approx N \log N - N \qquad (A.31)$$

We have the final result for the SMI of $N$ indistinguishable particles in a box of volume $V$, and temperature $T$:

$$H(1,2,\dots N) = N \log \left[\frac{V}{N} \left(\frac{2\pi m k_B T}{h^2}\right)^{3/2}\right] + \frac{5}{2} N \qquad (A.32)$$

By multiplying the SMI of $N$ particles in a box of volume $V$ at temperature $T$, by the factor ($k_B \log_e 2$), one gets the *entropy*, the *thermodynamic entropy* of an ideal gas of simple particles. This equation was derived by Sackur and by Tetrode in 1912, by using the Boltzmann definition of entropy.

One can convert this expression into the entropy function $S(E, V, N)$, by using the relationship between the total energy of the system, and the total kinetic energy of all the particles:

$$E = N \frac{m\langle v \rangle^2}{2} = \frac{3}{2} N k_B T \qquad (A.33)$$

The explicit entropy function of an ideal gas is:

$$S(E, V, N) = N k_B \ln \left[\frac{V}{N} \left(\frac{E}{N}\right)^{3/2}\right] + \frac{3}{2} k_B N \left[\frac{5}{3} + \ln \left(\frac{4\pi m}{3h^2}\right)\right] \qquad (A.34)$$

*(ii) Correlation Due to Intermolecular Interactions*

In Equation (A.34) we got the entropy of a system of non-interacting simple particles (ideal gas). In any real system of particles, there are some interactions between the particles. Without getting into any details on the function $U(r)$, it is clear that there are two regions of distances $0 \leq r \lesssim \sigma$ and $0 \leq r \lesssim \infty$, where the slope of the function $U(r)$ is negative and positive, respectively. Negative slope correspond to repulsive forces between the pair of the particles when they are at a distance smaller than $\sigma$. This is the reason why $\sigma$ is sometimes referred to as the *effective diameter* of the particles. For larger distances, $r \gtrsim \sigma$ we observe attractive forces between the particles.

Intuitively, it is clear that interactions between the particles induce *correlations* between the locational probabilities of the two particles. For hard-spheres particles there is infinitely strong repulsive force between two particles when they approach to a distance of $r \leq \sigma$. Thus, if we know the location $\boldsymbol{R_1}$ of one particle, we can be sure that a second particle, at $\boldsymbol{R_2}$ is not in a sphere of diameter $\sigma$ around the point $\boldsymbol{R_1}$. This *repulsive* interaction may be said to introduce *negative correlation* between the locations of the two particles.

On the other hand, two argon atoms *attract* each other at distances $r \lesssim 4\text{Å}$. Therefore, if we know the location of one particle say, at $\boldsymbol{R_1}$, the probability of observing a second particle at $\boldsymbol{R_2}$ is *larger* than the probability of finding the particle at $\boldsymbol{R_2}$ in the absence of a particle at $\boldsymbol{R_1}$. In this case we get *positive correlation* between the locations of the two particles.

We can conclude that in both cases (attraction and repulsion) there are correlations between the particles. These correlations can be cast in the form of *mutual information* which reduces the SMI of a system of $N$ simple particles in an ideal gas. The mathematical details of these correlations are discussed ref:3-5.

Here, we show only the form of the mutual information at very low density. At this limit, we can assume that there are only *pair correlations*, and neglect all higher order correlations. The mutual information due to these correlations is:

$$I(due\ to\ correlations\ in\ pairs)$$

$$= \frac{N(N-1)}{2} \int p(\boldsymbol{R_1}, \boldsymbol{R_2}) \log g(\boldsymbol{R_1}, \boldsymbol{R_2}) d\boldsymbol{R_1} d\boldsymbol{R_2} \qquad (A.35)$$

where $\boldsymbol{g}(R_1, R_2)$ is defined by:

$$g(\boldsymbol{R_1}, \boldsymbol{R_2}) = \frac{p(\boldsymbol{R_1}, \boldsymbol{R_2})}{p(\boldsymbol{R_1})p(\boldsymbol{R_2})} \qquad (A.36)$$

Note again that log g can be either positive or negative, but the average in (A.36) must be positive

## A.6 Conclusion

We summarize the main steps leading from the SMI to the entropy. We started with the SMI associated with the *locations* and *momenta* of the particles. We calculated the distribution of the locations and momenta that *maximizes* the SMI. We referred to this distribution as the *equilibrium distribution*. Let us denote this distribution of the locations and momenta of all the particles by $f_{eq}(\boldsymbol{R}, \boldsymbol{p})$.

Next, we use the equilibrium distribution to calculate the SMI of a system of $N$ particles in a volume $V$, and at temperature $T$. This SMI is, up to a multiplicative constant ($k_B \ln 2$) identical with the *entropy* of an ideal gas at equilibrium. This is the reason we referred to the distribution which maximizes the SMI as the *equilibrium distribution*.

It should be noted that in the derivation of the entropy, we used the SMI twice; first, to calculate the distribution that maximize the SMI, then evaluating the maximum SMI corresponding to this distribution. The distinction between the concepts of SMI and entropy is essential. Referring to SMI (as many do) as entropy, inevitably leads to such an awkward statement: the maximal value of the entropy (meaning the SMI) is the entropy (meaning the thermodynamic entropy). The correct statement is that the SMI associated with locations and momenta is defined for any system; small or large, at equilibrium or far from equilibrium. This SMI, not the entropy, evolves into a maximum value when the system reaches equilibrium. At this state, the SMI becomes proportional to the entropy of the system.

Since the entropy is a special case of a SMI, it follows that whatever interpretation one accepts for the SMI, it will be automatically applied to the concept of entropy. The most important conclusion is that entropy is not a function of time. Entropy does not change with time, and entropy does not have a tendency to increase.

We said that the SMI may be defined for a system with any number of particles including the case $N = 1$. This is true for the SMI. When we talk about the entropy of a system we require that the system be very large. The reason is that only for such systems the entropy-formulation of the Second Law of thermodynamic is valid.

## Appendix B. The main assumption of Non-equilibrium Thermodynamics

Non-equilibrium thermodynamics is founded on the assumption of *local equilibrium*[32-35] . While it is true that this assumption leads to the entire theory of thermodynamics of non-equilibrium process, it is far from clear that the very assumption of local equilibrium can be justified.

Most textbooks on non-equilibrium thermodynamics starts with the reasonable assumption that in such system the intensive variables such as temperature $T$, pressure $P$, and chemical potential $\mu$ may be defined operationally in each small element of volume $dV$ of the system. Thus, one writes

$$T(\boldsymbol{R}, t), P(\boldsymbol{R}, t), \mu(\boldsymbol{R}, t) \qquad \text{(B.1)}$$

where $\boldsymbol{R}$ is the locational vector of a point in the system, and $t$ is the time. One can further assume that the density $\rho(\boldsymbol{R}, t)$ is defined locally at point $\boldsymbol{R}$ and integrate over the entire volume to obtain the total number of particles $N$

$$N = \int \rho(\boldsymbol{R}, t) d\boldsymbol{R} \qquad \text{(B.2)}$$

Similarly, one can define densities $\rho_R(\boldsymbol{R}, t)$ for each component of the system.

One also defines the local internal energy per unit of volume $u(\boldsymbol{R}, t)$. It is not clear however, how to integrate $u(\boldsymbol{R}, t)$ over the entire volume of the system to obtain the total internal energy of the system. While this may be done exactly for ideal gases, i.e. assuming that the total energy of the system is the sum of all the kinetic (as well as internal) energies of all the particles in the system, it is not clear how to do the same for systems having interacting particles. For suppose we divide the total volume of the system into $c$ small cells, and assume that the internal energy in cell $i$ is $u(i, t)$. In which case the total internal energy of the system is written as

$$U = \sum_i u(i, t) \qquad \text{(B.2)}$$

And the corresponding integral is

$$U = \int u(\boldsymbol{R}, t) d\boldsymbol{R} \qquad \text{(B.4)}$$

Here, the integration is essentially the sum of the $u(\boldsymbol{R}, t)$ in small cells in the system, neglecting the interaction energies between the different cells. If there are interactions among the particles, then the internal energy of the system cannot be written as a sum of the form (B.3) or (B.4).

The most important and unjustified assumption is it is related to the local entropy $s(\boldsymbol{R}, t)$. One assumes that the entropy function, say $S(U, V, N)$ is the same function for the local entropy, i.e. $s$ is the same function as the local energy, volume, and number of particles of each element of volume.

Thus, the local entropy of the cell $i$ is presumed to be the same function of the local energy, volume, and number of particles, i.e. one writes

$$S = \sum_i s(i, t) = \int_V s(\boldsymbol{R}, t) d\boldsymbol{R} \qquad \text{(B.5)}$$

This assumption may be justified for ideal gas when the distribution of locations and velocities is meaningful for each element of volume in the system. To the best of the author's knowledge this assumption has never been justified for systems of interacting particles. The main problem in these definitions is that it does not take into account the correlations between different cells or between different elements of volume in the continuous case. These correlations depend on the number of particles in each cell which changes with time.

Once one makes such an assumption one can write the changes in the entropy of the system as

$$dS = d_e S + d_i S \qquad (B.6)$$

where $d_e S$ is the entropy change due to the heat exchange between the system and its surrounding, and $d_i S$ is the entropy produced in the system. For an isolated system $d_e S$, and all the entropy change is due to $d_i S$. The latter is further written as

$$\frac{d_i S}{dt} = \int \sigma d\mathbf{R} \qquad (B.7)$$

where $\sigma(\mathbf{R}, t)$ is the *local* entropy production

$$\sigma(\mathbf{R}, t) = \frac{d_i S}{dt} \geq 0 \qquad (B.8)$$

Thus, for an isolated system one has a local entropy production which is a function of time, and after integration one obtains also the total change of the entropy of the system as a function of time. Since the quantity $\sigma$ is defined in terms of the local entropy function $s(\mathbf{R}, t)$, and since $s(\mathbf{R}, t)$ is not a well-defined quantity, one should doubt the whole theory based on the assumption of local equilibrium.

**References**

1. Atkins, P. (1984), *The Second Law*, Scientific American Books, W. H. Freeman and Co., New York

2. Atkins, P. (2007), *Four Laws That Drive The Universe*, Oxford University Press

3. Ben-Naim, A. (2016), *Entropy the Truth the Whole Truth and Nothing but the Truth*, World Scientific Publishing, Singapore

4. Ben-Naim, A. (2017), *The Four Laws that do not drive the Universe*, World Scientific Publishing, Singapore

5. Ben-Naim A. and Casadei D. (2017*), Modern Thermodynamics*, World Scientific Publishing, Singapore

6. Shannon, C. E. (1948), *A Mathematical Theory of Communication*, Bell System Tech. J., 27

7. Ben-Naim, A. (2009), *An Informational-Theoretical Formulation of the Second Law of Thermodynamics*, J. Chem. Education, 86, 99.

8. Ben-Naim, A. (2012), *Entropy and the Second Law, Interpretation and Misss Interpretationsss*, World Scientific Publishing, Singapore.

9. Brush, S. G. (1976), *The Kind Of Motion We Call Heat. A History Of The Kinetic Theory of GasesIn The 19th Century*, *Book 2: Statistical Physics and Irreversible Processes*. North-Holland Publishing Company.

10. Brush, S. G. (1983), *Statistical Physics and the Atomic Theory of Matter, from Boyle and Newton to Landau and Onsager*. Princeton University Press, Princeton.

11. Lemons, D. S. (2013), *A student's Guide to Entropy*. Cambridge University Press.

12. Boltzmann, L. (1896), *Lectures on Gas Theory*, Translated by S.G. Brush, Dover, New York (1995)

13. Gibbs, J.W. (1906), *Collected Scientific Papers of J. Willard Gibbs*, Longmans Green, New York

14. Goldsein, S. (2001), *Boltzmann's Approach to Statistical Mechanics*. (Published in arXiv:cond-mat/0105242,v1, 11 May 2001).

15. Jaynes, E.T. (1965), *Gibbs vs Boltzmann Entropies*, *Am. J. Phys.*, 33, 391

16. Lebowitz, J.L. (1993), *Boltzmann's Entropy and Time's Arrow*, Physica Today, 46, 32

17. Wehrl, A. (1991), *The Many Faces of Entropy,* Reports on Mathematical Physics, 30, 119

18. Hill, T.T. (1960), An Introduction to Statistical Thermodynamics, Addison-Wesley Publishing Company, Inc. Reading Mass. USA.

19. Mackey, M. C. (1992), *Time's Arrow, The Origins of Thermodynamic Behavior*, Dover Publications, New York

20. Ben-Naim, A. (2008), *A Farewell to Entropy: Statistical Thermodynamics Based on Information*, World Scientific Publishing, Singapore

21. Jaynes, E.T. (1957), *Information Theory and Statistical Mechanics*, Phys.Rev., 106, 620

22. Jaynes, E.T. (1957), *Information Theory and Statistical Mechanics II*, Phys. Rev., 108, 171

23. Jaynes, E.T. (1973), *The well-posed Problem*, Chapter 8 in Jaynes (1983)

24. Jaynes, E.T. (1983), *Papers on Probability, Statistics, and Statistical Physics*,

Edited by R.D. Rosenkrantz, D. Reidel Publishing Co., London

25. Ben-Naim, A. (2017), *Entropy, Shannon's Measure of Information and Boltzmann's H-Theorem,in Entropy*, 19, 48-66, (2017).

26. Sackur, O. (1911), *Annalen der Physik*, 36, 958

27. Tetrode, H. (1912), *Annalen der Physik*, 38, 434

28. Ben-Naim, A. (2017), *Information Theory, Part II: Introduction to the Fundamental Concepts,* World Scientific, Singapore.

29. Ben-Naim, A. (2015), *Information, Entropy, Life and the Universe, What We Know and What We Do Not Know*, World Scientific Publishing, Singapore

30. Penrose, R. (1989), *The Emperor's Mind. Concerning Computers, Minds and the Law of Physics,* Penguin Books, New York

31. Carroll, S. (2010), *From Eternity to Here, The Quest for the Ultimate Theory of Time*, Plume, USA

32. de Groot, S. R., and Mazur, P. (1962), *Non-Equilibrium Thermodynamics*, North-Holland Publishing Co., Amsterdam

33. Kondepudi, D. and Prigogine, I. (1998), *Modern Thermodynamics From Heat Engines to Dissipative Structures*, John Wiley and Sons, England

34. Sethna, J. P. (2006), *Statistical Mechanics: Entropy, Order Parameters and Complexity*, Oxford University Press, Oxford

35. Kreuzer, H. J. (1981), *Non-equilibrium Thermodynamics and Statistical Foundations*, Oxford University Press, Oxford

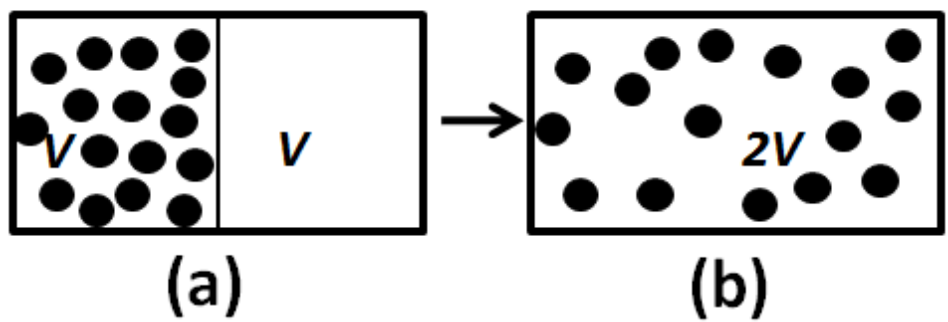

**Figure 1. Expansion of an ideal gas from *V* to 2 *V***

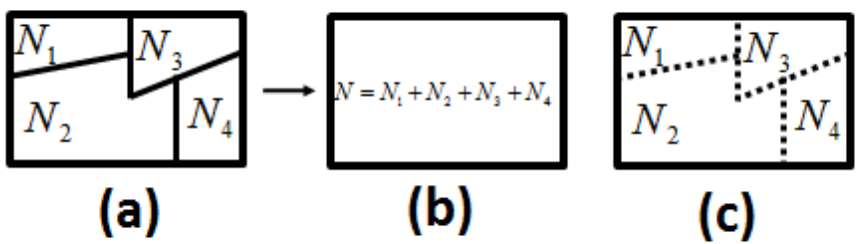

**Figure 2. (a) A constrained equilibrium system.**
**(b) The unconstrained equilibrium state**
**(c) The same system as in**
**(b) (unconstrained) but with a distribution of particles as in (a)**

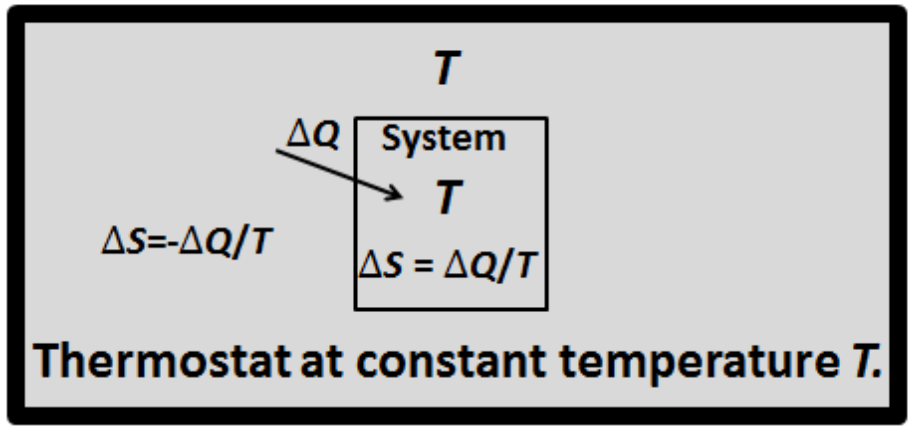

Figure 3 A system in contact with an isolated heat reservoir
( a thermostat).

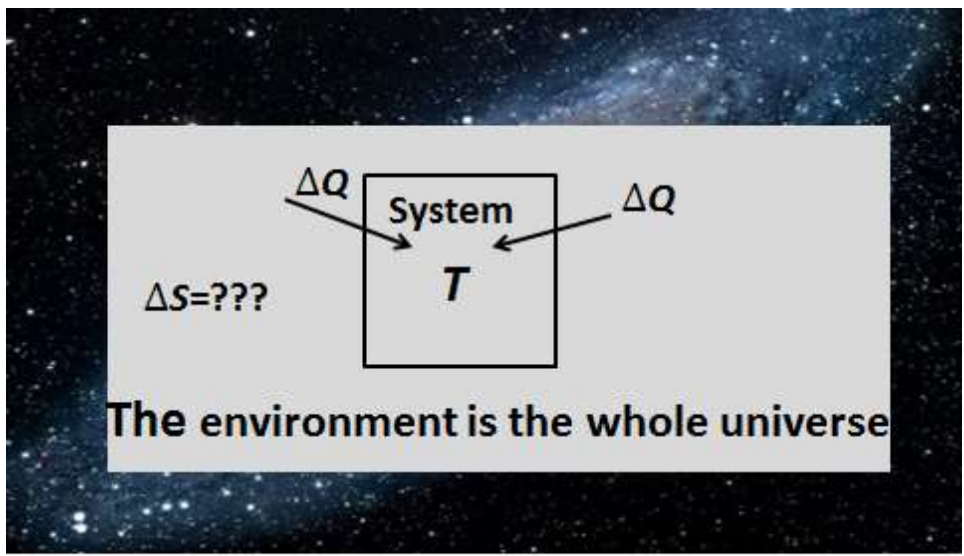

Figure 4. A system in contact with the entire universe

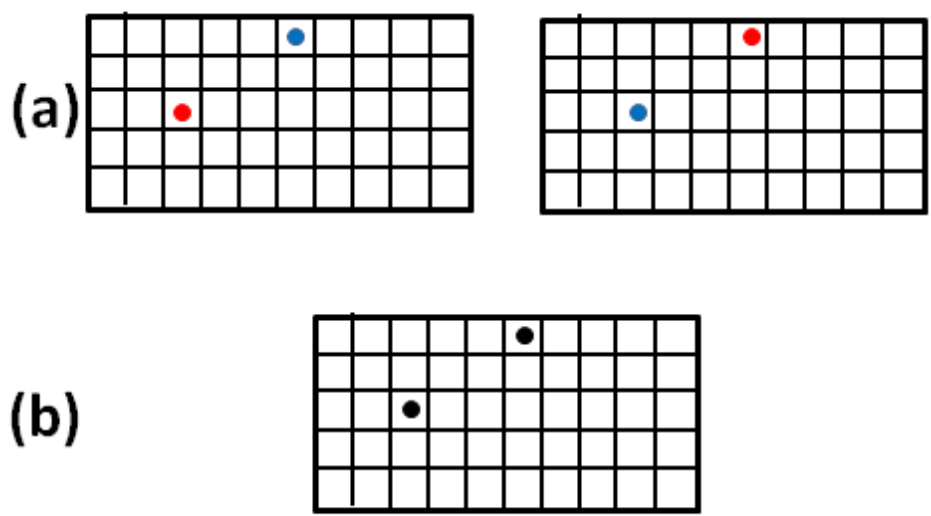

**Figure 5** (a) Two different configurations become
(b) identical when the particles are indistinguishable